\documentclass{optica-article}

\journal{opticajournal} 

\articletype{Research Article}

\usepackage{lineno}

\begin{document}

\title{Single-Crystal, Single-Chirality, Single-Wall Carbon Nanotube Heterostructures for Optoelectronics: An Opinion}

\author{Ting-Wei Chang,\authormark{1,2} Gustavo M.\ Rodriguez-Barrios,\authormark{1,2}\\ Andrey Baydin\authormark{1,3,4} and Junichiro Kono\authormark{1,3,4,5,6,*}}

\address{\authormark{1}Department of Electrical and Computer Engineering, Rice University, 6100 Main Street, Houston, TX 77005, USA\\
\authormark{2}Applied Physics Graduate Program, Smalley--Curl Institute, Rice University, 6100 Main Street, Houston, TX 77005, USA\\
\authormark{3}Smalley--Curl Institute, Rice University, 6100 Main Street, Houston, TX 77005, USA\\
\authormark{4}{Rice Advanced Materials Institute, Rice University, Houston, Texas 77005, USA}
\authormark{5}Department of Physics and Astronomy, Rice University, 6100 Main Street, Houston, TX 77005, USA\\
\authormark{6}Department of Materials Science and NanoEngineering, Rice University, 6100 Main Street, Houston, TX 77005, USA}

\email{\authormark{*}kono@rice.edu} 


\begin{abstract}
The extraordinary one-dimensional properties of carbon nanotubes have captivated scientists and engineers since their discovery in the early 1990s. In particular, semiconducting single-wall carbon nanotubes (SWCNTs) are highly promising for optoelectronic applications because of their diameter-dependent direct band gaps and strong, tunable light--matter interactions. However, the prevalence of structural disorder, misalignment, and chirality heterogeneity in macroscopic assemblies has hindered their practical applications. Recently, advanced assembly methods, combined with post-growth chirality separation techniques, have enabled the fabrication of wafer-scale, nearly crystalline films of highly aligned and densely packed SWCNTs with tailored properties. In this Opinion, we discuss how these films provide a transformative platform for engineering ``Single$^3$'' heterostructures—assemblies that are simultaneously \emph{single}-crystal, \emph{single}-chirality, and \emph{single}-wall. Stacking these layers with nanometer-scale precision and tunable thicknesses allows for the realization of artificial bilayer junctions, quantum wells, and superlattices. We posit that these architectures will enable a new generation of high-performance devices, including lasers, photodiodes, solar cells, and single-photon emitters. 
\end{abstract}


\section{Introduction}
Single-wall carbon nanotubes (SWCNTs) combine tunable energy band gaps, ultrahigh carrier mobilities, and strong excitonic effects arising from their one-dimensional (1D) electronic structure~\cite{Aurus2008,Dresselhaus2001,Dresselhaus1998}. Their optical transitions, such as $E_{11}$ and $E_{22}$, span from the near-infrared to the visible spectrum, making them promising candidates for photodetectors, light emitters, modulators, and quantum photonic devices~\cite{doumani_macroscopically_2025,gao_science_2019,baydin_carbon_2022,wang_advancement_2024,nanot_optoelectronic_2012,XieEtAl2025NRP,samahaGrapheneTerahertzDevices2024}. However, despite decades of research, the practical implementation of SWCNTs as functional semiconductor materials has been hindered by the inability to assemble them into macroscopic structures with long-range order and electronic uniformity.
 
Traditional solution-processed SWCNT films consist of randomly oriented nanotube networks, where electronic transport is dominated by tube-to-tube junction resistance and percolation effects~\cite{cao_ultrathin_2009}. Orientational disorder obscures the unique intrinsic 1D properties of individual nanotubes, while mixed chiralities lead to uncontrolled band gap heterogeneity. These factors broaden optical linewidths and reduce device reproducibility. Although post-growth chirality separation techniques such as aqueous two-phase extraction (ATPE)~\cite{fagan_aqueous_2019}, polymer-assisted sorting~\cite{hata_water-assisted_2004}, and density-gradient ultracentrifugation (DGU)~\cite{arnold_sorting_2006} have improved access to single-chirality nanotubes, material purity alone is insufficient. Realizing their full potential requires ordered ensemble arrays of aligned SWCNTs with precisely controlled density, alignment, and chirality purity.

Several methods have been explored to align SWCNTs. DNA-templated assembly and molecular recognition strategies guide nanotubes with high selectivity, but the resulting films are usually sparse and difficult to scale to wafer-level areas~\cite{maune_self-assembly_2010, li_dna-directed_2005}. Dimension-limited self-alignment leverages nanoscale confinement or topographic templates to orient nanotubes; however, achieving high packing density remains challenging due to template geometry~\cite{chao_small_2024,komatsu_groove-assisted_2020}. Chemical vapor deposition (CVD) on crystalline substrates, such as quartz, can produce highly aligned arrays with excellent electronic properties~\cite{ismach_direct_2010}, but faces challenges in controlling chirality and transferring the arrays to arbitrary substrates. Lastly, Langmuir--Blodgett assembly offers relatively uniform aligned monolayers~\cite{li_langmuirblodgett_2007}, but scaling to large areas while stacking multilayers remains nontrivial.

\begin{figure}[htbp]
\centering\includegraphics[width=15cm]{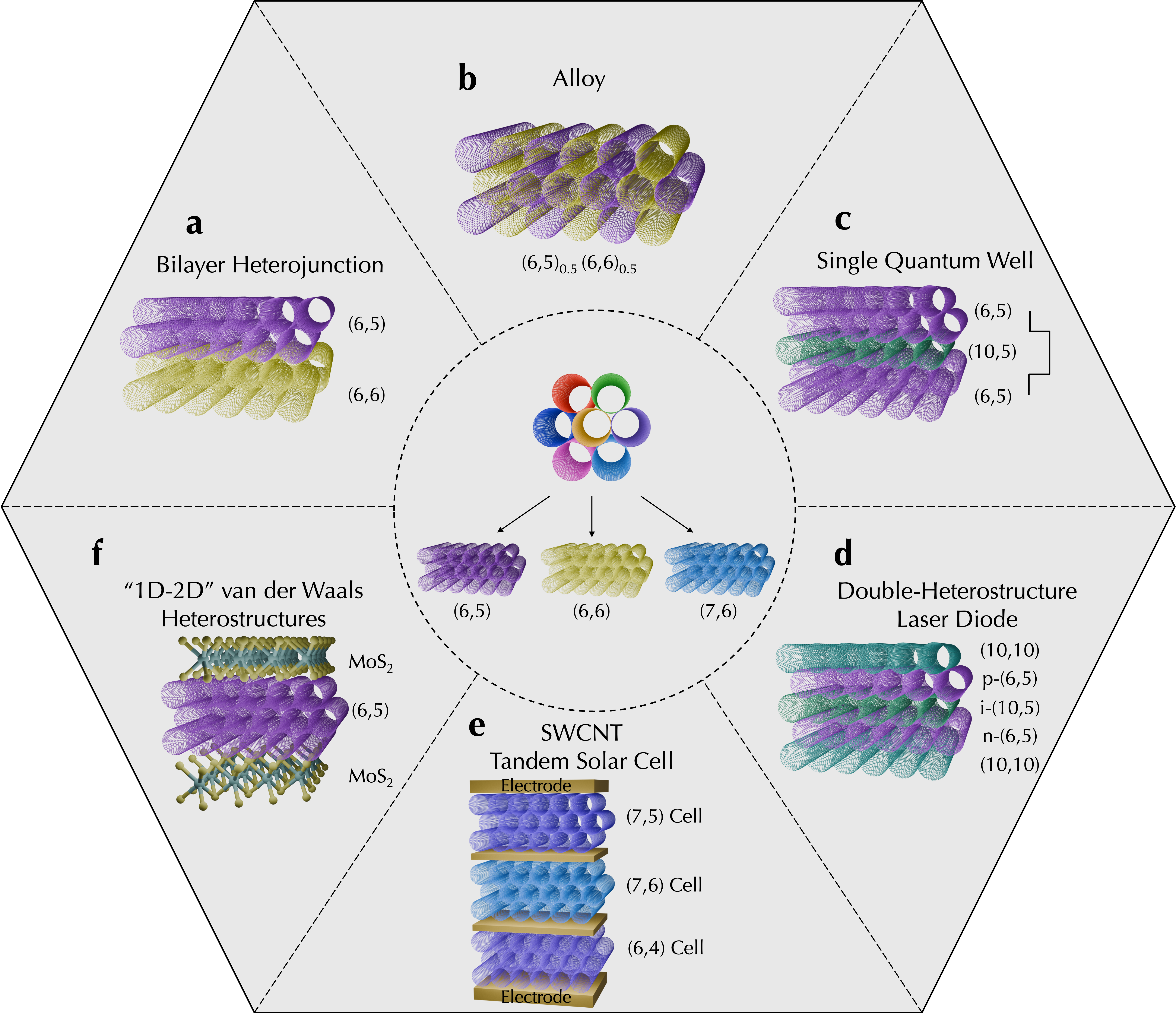}
\caption{Design framework for 1D SWCNT heterostructure engineering. At the center, chirality-defined, highly aligned SWCNT films serve as the fundamental building blocks. Where each chirality, such as (6,5), (6,6), (7,6), provides a distinct bandgap and excitonic transition. Surrounding the center, representative configurations are illustrated. (a)~Bilayer heterojunction formed by stacking two chirality-pure SWCNT films. (b)~Chirality alloy films composed of mixed SWCNT species by co-filtering multiple chiralities. (c)~Single quantum well structures, where a narrow-bandgap SWCNT layer is sandwiched between wide-bandgap layers. (d)~Double heterostructure laser diode architectures, in which carrier injection and recombination are engineered through multiple chirality-defined interfaces. (e)~SWCNT tandem solar cell, where multiple absorber layers with different bandgaps are vertically integrated. (f)~1D–2D van der Waals heterostructure that combine SWCNT films with 2D materials such as MoS$_2$.}
\label{fig:omex}
\end{figure}

Controlled vacuum filtration (CVF) addresses these challenges by enabling wafer-scale SWCNT films with high packing density, deterministic thickness, and nematic alignment (order parameter $S>0.9$)~\cite{he_wafer-scale_2016}. These aligned films exhibit strong optical anisotropy and consistent macroscopic transport properties, establishing a materials platform where SWCNTs function as crystalline 1D semiconductor solids rather than disordered networks. By using CVF films as modular building blocks~\cite{DoumaniEtAl2023NC,Rodriguez-BarriosEtAl2025}, one can engineer SWCNT heterostructures with tailored properties through precise stacking and composition; see Fig.~\ref{fig:omex}.

\section{SWCNT Heterostructures}
A key feature of this heterostructure platform is the bilayer heterojunction, created by stacking two chirality-pure SWCNT films with different band gaps, such as $(6,5)/(6,6)$ pairing [Fig.~\ref{fig:omex}(a)]. Depending on the chirality selection, these interfaces can exhibit type-I or type-II band alignments, enabling phenomena like carrier confinement, exciton funneling, and charge separation without chemical doping. Unlike conventional CNT junctions relying on electrostatic gating or chemical modifications~\cite{qian_band_2020,kawasaki_complex_2024,daneshvar_critical_2021}, bilayers defined by chirality can achieve intrinsic band engineering while preserving structural order and excitonic coherence.  

Beyond simple interfaces, we can engineer SWCNT films by co-filtering multiple chiralities to create compositional alloys that are statistically mixed but structurally ordered, as illustrated in [Fig.~\ref{fig:omex}(b)]. Such films enable continuous tuning of the effective band structure and density of states, analogous to alloy engineering in III–V semiconductors. The resulting chirality alloys could lead to enhanced broadband absorption layers and graded energy landscapes essential for multispectral photodetectors and photovoltaic devices, all while preserving the long-range alignment and anisotropic transport properties absent in conventional mixed CNT networks. 

By alternating wide- and narrow-bandgap SWCNTs, we can create quantum wells and superlattices [Fig.~\ref{fig:omex}(c)]. For instance, stacking a $(6,5)/(10,5)/(6,5)$ structure forms a confined potential well capable of trapping excitons or carriers. When periodically repeated, such stacks enable miniband formation and resonant tunneling phenomena~\cite{perebeinos_exciton_2007}. Based on established exciton physics in SWCNTs~\cite{dresselhaus_exciton_2007}, these structures are now feasible through ordered, stackable single-chirality SWCNT films. 

Extending these designs into active devices enables an all-carbon double-heterostructure laser diode [Fig.~\ref{fig:omex}(d)]. In this architecture, a layer of narrow-bandgap SWCNTs serves as the active region, sandwiched between higher-bandgap cladding layers. This design mirrors the principles of III–V double-heterojunction lasers, leveraging the strong excitonic oscillator strength and broad tunability intrinsic to SWCNTs. 

Furthermore, these concepts facilitate the design of complex optoelectronic device architectures. For example, SWCNT tandem solar cells can be realized by vertically stacking multiple chirality-selective absorber layers [Fig.~\ref{fig:omex}(e)]. In contrast to percolative CNT photovoltaics, these stacked architectures enable thickness optimization and controlled carrier extraction, much like in multijunction III–V solar cells.

Beyond purely nanotube structures, integrating CVF-aligned SWCNT films with 2D materials can yield mixed-dimensional van der Waals heterostructures [Fig.~\ref{fig:omex}(f)]. Stacking SWCNTs with MoS$_2$, hBN, or graphene enables charge transfer, control dielectric screening, and generates hybridized excitonic states. These 1D–2D heterostructures extend van der Waals engineering, opening new opportunities for exciton manipulation and anisotropic light–matter coupling. 

Together, these examples demonstrate that highly aligned, densely packed, single-chirality SWCNTs can serve as programmable 1D semiconductor building blocks. By controlling chirality, thickness, and stacking order, one can achieve band-structure and excitonic engineering at the material level, mirroring the sophistication of III–V heterostructures while enhancing excitonic interactions and mechanical flexibility. This platform provides a foundation for photovoltaics, light emitters, and quantum photonic devices.

\section{Challenges and Outlook}
While CVF-enabled SWCNT heterostructures offer significant potential, several challenges remain before they can be fully translated into scalable optoelectronic technologies. One major challenge is that, although chirality purification has improved significantly, current methods still yield mixtures that require extensive processing~\cite{doumani_macroscopically_2025}. It remains difficult to produce gram quantities of single-chirality SWCNTs. This limits the use of large-area quantum wells and double heterostructures. [Fig.~\ref{fig:omex}(c,e)]. Continued improvements in high-throughput sorting and selective synthesis are vital to enable complex stacks with consistent properties~\cite{kharlamova_synthesis_2022}.

Another factor is interfacial quality. Residual surfactants and processing contaminants can trap excitons and persist after filtration. These residues impact optical dynamics and charge transport~\cite{christensen_localized_2022,ostos_potentiometric_2021} and are especially detrimental in structures relying on coherent excitonic interactions, such as bilayer heterojunctions and mixed-dimensional structures [Fig.~\ref{fig:omex}(a,e)]. Developing gentle transfer methods, surfactant-free dispersion strategies, and \textit{in~situ} cleaning approaches is essential for achieving high-quality interfaces~\cite{kim_recent_2023}.

Additionally, reliable control over carrier type and density is critical for device architectures as tandem solar cells and laser diodes [Fig.~\ref{fig:omex}(d,f)]. While chemical dopants can induce $p$-type or $n$-type behavior, these states often lack thermal stability and environmental robustness~\cite{kawasaki_complex_2024}. Electrostatic gating and charge-transfer doping offer promising pathways, but achieving uniform, stable doping in multilayer SWCNT stacks remains challenging~\cite{xu_efficient_2017}.

Unlike traditional III–V systems, SWCNT heterostructures require a new predictive model. The anisotropic coupling between tubes~\cite{crochet_electrodynamic_2011,simpson_resonance_2018}, strong excitonic effects~\cite{jiang_chirality_2007,walsh_screening_2007}, and chirality-dependent dielectric screening~\cite{nugraha_dielectric_2010,araujo_diameter_2009} demand theoretical and computational frameworks to design quantum wells and mixed-dimensional interfaces effectively.

Finally, achieving low-resistance contacts remains a bottleneck~\cite{cao_ultrathin_2009,he_wafer-scale_2016}. However, the potential of aligned SWCNT films to be solution-processed provides a clear integration path for CMOS electronics and flexible photonic systems~\cite{cao_arrays_2013}.

\section{Conclusion}

In summary, CVF has transformed the assembly of SWCNTs into solid-state materials. Disordered networks are replaced by highly aligned chirality-defined thin films that function as crystalline 1D semiconductor solids. This development enables programmable SWCNT heterostructures with precisely tunable composition and stacking, enabling band alignment engineering. This supports a variety of configurations, from double-heterojunction laser diodes to tandem solar cells. Overall, this Opinion establishes a roadmap for bringing concepts that have long defined III–V optoelectronics into the ``1D regime,'' facilitating optoelectronic and quantum photonic devices that are unattainable in traditional semiconductors. 

\bibliography{references}

\end{document}